\documentclass[11pt]{article}
\usepackage[utf8]{inputenc}
\usepackage[margin=1.3in]{geometry}
\usepackage{afterpage}

\usepackage{cite}
\usepackage[table,xcdraw]{xcolor}
\usepackage{amsmath}
\usepackage{stmaryrd}

\usepackage{amsthm}
\usepackage{tikz}
\usetikzlibrary{positioning}
\usetikzlibrary{cd}
\usetikzlibrary{decorations.markings}
\usepackage{pgfplots}
\usepackage[all]{xy}
\theoremstyle{definition}

\newtheorem{theorem}{Theorem}[section]

\usepackage{amsthm}

\usepackage{epigraph}
\usepackage[T1]{fontenc}
\usepackage{kpfonts,baskervald}
\usepackage{lipsum}
\usepackage{hyperref}
\hypersetup{
    colorlinks,
    citecolor=blue,
    filecolor=black,
    linkcolor=blue,
    urlcolor=blue
}

\numberwithin{equation}{section}
\usepackage[labelfont=bf]{caption}
\usepackage{blindtext}
\usepackage{braket}
\DeclareMathOperator{\Tor}{Tor}

\newcommand{\df}{\coloneqq}

\usepackage{subfiles}
\begin{document}
\titlepage
\medskip
\phantom{ghost}

\vspace{2cm}

\begin{center}
    {\LARGE {\bf{Maruyoshi-Song Flows and Defect Groups of $D_p^b(G)$ Theories}}}
\end{center}

\vspace{1.5cm}

\begin{center}
    {Saghar S. Hosseini$^*$ and Robert Moscrop$^\dagger$}\end{center}

\begin{center}
{\small $*$ Department of Mathematical Sciences and Centre for Particle Theory \\ Durham University, Durham, DH1 3LE, UK}\\[0.15in]

{\small $\dagger$ Department of Mathematics \\
Uppsala University, Uppsala, Sweden}
\end{center}

\vspace{1.2cm}

\begin{abstract}
    \noindent We study the defect groups of $D_p^b(G)$ theories using geometric engineering and BPS quivers. In the simple case when $b=h^\vee (G)$, we use the BPS quivers of the theory to see that the defect group is compatible with a known Maruyoshi-Song flow. To extend to the case where $b\neq h^\vee (G)$, we use a similar Maruyoshi-Song flow to conjecture that the defect groups of $D_p^b(G)$ theories are given by those of $G^{(b)}[k]$ theories. In the cases of $G=A_n, \;E_6, \;E_8$  we cross check our result by calculating the BPS quivers of the $G^{(b)}[k]$ theories and looking at the cokernel of their intersection matrix.  
\end{abstract}

\vspace{5.0cm}

\noindent $\underline{\qquad\qquad\qquad\qquad\qquad\qquad\qquad}$\\
\noindent {\footnotesize June 2021}

\vspace{0.2cm}

\noindent Emails: {\footnotesize{\tt sagharsadat.hosseinisemnani@durham.ac.uk}, {\tt robert.moscrop@math.uu.se}}


\newpage
{\hypersetup{linkcolor=black}
\tableofcontents
}

\section{Introduction}
Recently, there has been much interest in understanding the spectrum of extended operators such as Wilson lines and surface defects in field theories \cite{Gaiotto:2014kfa, Kapustin:2014gua, Gukov:2013zka, Kapustin:2013uxa, Gaiotto:2010be, BenettiGenolini:2020doj}. In fact, there are theories with the same local structure that differ only by their spectrum of extended operators \cite{Aharony:2013hda} and consequently by their partition functions, further highlighting the importance of their presence. Furthermore, the charged extended operators in the theory have associated global
symmetries known as {\it higher form symmetries}. 

Originally, higher form symmetries were obtained by looking at the gauge theory description of field theories. For example, in four dimensional Yang-Mills theory with a fixed gauge algebra there are multiple choices of a gauge group. Each choice of gauge group corresponds to a distinct spectrum of line operators each with their own associated higher form symmetry. Therefore, one may learn about the global structure of the theory by studying its gauge theory description. However, not all theories have a known weakly-coupled gauge theory description and a corresponding gauge group. As such, it is natural to look for an alternative method to study higher form symmetries. In fact, for a large class of theories the classification of possible global structures naturally arises from constructing QFTs from string theory \cite{Garcia-Etxebarria:2019cnb}.

In this paper, we are interested in understanding the higher form symmetries of a large class of $\mathcal{N}=2$ 4d SCFTs known as $G^{(b)}[k]$ and $D_p^b(G)$ \cite{dpg1,dpg2, Wang:2015mra, Closset:2020afy, Carta:2021whq,Giacomelli:2017ckh,Xie:2012hs}. These theories living on $M_4$ are geometrically engineered by considering type IIB string theory on $M_4\times \mathcal{W}_6$, where $\mathcal{W}_6$ is a non-compact space with an isolated hypersurface singularity. 

As studied in \cite{Garcia-Etxebarria:2019cnb}, given the above type of the geometry, we expect some of these 4d theories to have 1-form symmetries. In fact, just as in 4d Yang-Mills, for a theory with a fixed local structure, there are multiple choices of a global structure and so a 1-form symmetry. For example, for 4d Yang-Mills theory we can either have a magnetic 1-form symmetry, an electric 1-form symmetry or a dyonic one.
To understand why this is, one can consider the stringy origin of the 4d theories.
When we embed a 4d theory in type IIB string theory, the generators of the 1-form symmetries in 4d have a counterpart in the IIB theory, namely the RR flux operators. It is known that (in the presence of torsion) the RR flux operators in 10d do not commute with each other; in other words, these flux operators are mutually non-local \cite{Witten:1998wy, Freed:2006ya, Freed:2006yc}. Furthermore, as we will see in the next sections, when we do dimensional reduction to 4d, only a \emph{commuting} subset of the flux operators in 10d can be realised in 4d to give rise to the operators generating the 1-form symmetries in 4d. Thus, for each choice of a commuting set of operators in 10d we have a different corresponding choice of a 1-form symmetry in our 4d theory.

From a field theory perspective, the origin of the non-commutativity of flux operators can be understood by analysing the mutual locality of the line operators in the theory by expanding on the ideas of \cite{Aharony:2013hda}. In fact, the later picture was generalised to non-Lagrangian CFTs in \cite{DelZotto:2015isa}. 

The relevant algebraic object to study here is called the defect group. More specifically, line or surface operators, arising from branes wrapping non-compact cycles in the internal space $\mathcal{W}_6$, are charged under the higher form symmetry group. Such defects may be screened by the dynamical operators which arise from branes wrapping compact cycles. The non-compact cycles are elements of the homology of $\mathcal{W}_6$ relative to its boundary so they extend to the boundary of $\mathcal{W}_6$ at infinity. It is precisely for this reason that they result in defects which have infinite tension. The defect group measures the defects which are not screened by dynamical objects, and as we will see, it is defined by the (relative) homology of $\mathcal{W}_6$. It has been shown that this object is readily computable through several methods \cite{Morrison:2020ool, Albertini:2020mdx, Dierigl:2020myk, 4dpaper,   Bah:2020uev, Closset:2020scj,  Apruzzi:2020zot, Bhardwaj:2020phs, Gukov:2020btk,Bhardwaj:2021pfz, Apruzzi:2021phx, Apruzzi:2021vcu} giving us considerable insight into the higher form symmetries of various field theories.

Our aim here is to compute the defect groups of $G^{(b)}[k]$ and $D_p^b(G)$ theories. As we will explain, this is given by the cokernel of the Dirac pairing \cite{4dpaper}. This gives us two possible approaches, each with their own constraints. The first method relies on the application of Orlik's theorem \cite{OM} where the cokernel of $G^{(b)}[k]$ is directly calculated from the internal space. However, this theorem only applies to quasi-homogeneous polynomials in $\mathbb{C}^n$ and it is not immediately obvious whether it can be generalised and applied to the $D_p^b(G)$ theories whose internal geometry is embedded in $\mathbb{C}^\times\times\mathbb{C}^3$.

The second method is to inspect the BPS quivers of the theories, should one exist. This method does not care about the embedding space of the internal geometry, but one does not immediately find a closed form expression when looking at a series of singularities as one would using Orlik's algorithm.

Furthermore, the higher form symmetries are invariant under Maruyoshi-Song flows \cite{Maruyoshi:2016tqk, Maruyoshi:2016aim, Agarwal:2016pjo}. In our case of interest, there are flows connecting $G^{(b)}[k]$ and $D_p^b(G)$ as well as some of these to the generalised Argyres-Douglas theories known as $(G, G')$ theories. The latter theories were constructed also by putting type IIB theory on a special class of isolated hypersurface singularities \cite{Cecotti:2010fi, Xie:2012hs,  Xie:2015rpa, Xie:2016evu}. Therefore, we will use these flows to find the defect groups of $D_p^b(G)$ theories and verify our results from different perspectives.

The structure of the paper is as follows. In section 2 we briefly review the view of higher-form symmetries from the geometric engineering standpoint and highlight the two methods of computation we will use. Section 3 then deals with the case of $D_p(G)$ theories from which we conclude that the defect groups are compatible with an RG flow found in \cite{Giacomelli:2017ckh}. We then calculate the defect groups of $G^{(b)}[k]$ theories in section 4 and conjecture the $D_p^b(G)$ defect groups using the general form of the previously mentioned RG flow. Our results are presented in Table \ref{tbl:Gb(k)defects}. In order to verify these results, we also calculate the BPS quivers for the $J=A_n, E_6, E_8$ cases using the polar curve method \cite{gabrielov}\footnote{We thank Michele Del Zotto for suggesting this method to us.}.

\section{Higher form symmetries from geometric engineering}

\subsection{Geometric origin}
In this section we give a short review on how to find the discrete higher form symmetries of geometric engineered field theories following the methods developed in \cite{Garcia-Etxebarria:2019cnb}.

We may define a $4$-dimensional supersymmetric field theory, denoted by $\mathcal{T}_{\mathcal{W}_6}$, on $\mathcal{M}_4$ by putting type IIB string theory on
\begin{equation}
    \mathcal{M}_{10}=\mathcal{M}_4\times \mathcal{W}_6\, ,
\end{equation}
where $\mathcal{M}_4$ is chosen to be an orientable torsion-free manifold with a spin structure, and $\mathcal{W}_6$ is any manifold. In string theory, the $(p - k + 1)$-dimensional dynamical objects result from $p$-branes wrapping the compact $k$-cycles in the internal geometry and the $(p - k + 1)$-dimensional line and surface defects result from the $p$-branes wrapping non-compact $k$-cycles. Therefore, every $p$-brane wrapping a non-compact cycle could result in a defect group \cite{DelZotto:2015isa} for the discrete $(p - k + 1)$-form global symmetry.

Here we choose $\mathcal{W}_6$ to be a non-compact manifold. Let us emphasise that this is what results in interesting and non-trivial higher form symmetries. To see this, let us temporarily take $\mathcal{W}_6$ to be a compact manifold. Then, we do not have any defects since these are defined to be the objects wrapping the non-compact cycles and are only left with dynamical objects which break the 1-form symmetries. Thus, the defect group for the 1-form symmetry measuring the unscreened defects must be trivial. This is exactly in agreement with the fact that a theory coupled to gravity must not have any global symmetries \cite{ArkaniHamed:2006dz, Banks:2010zn}.

However, in a non-compact manifold we can have non-trivial torsional-cycles on the boundary. Then, the operators $\phi_\sigma$, with $\sigma$ a torsional class in the first K-theory group of the boundary $K^1(\partial \mathcal{M}_{10})$, measuring the torsional part of ﬂux expectation values for the IIB supergravity fields at infinity on a cycle Poincaré dual to $\sigma$ do not commute \cite{Freed:2006ya, Freed:2006yc}
\begin{equation}\label{noncommute}
    \phi_\sigma\phi_{\sigma'}=e^{2\pi i L(\sigma,\sigma')} \phi_{\sigma'}\phi_{\sigma},
\end{equation}
where $L$ is the linking pairing in $\partial\mathcal{M}_{10}$
\begin{equation}\label{linking}
    L: \Tor K^1(\partial \mathcal{M}_{10}) \times \Tor K^1(\partial \mathcal{M}_{10}) \to \mathbb{Q}/ \mathbb{Z}\, .
\end{equation}
Furthermore, as shown in \cite{4dpaper}, in our case the $K^1$ group of the boundary equals the direct sum of its odd cohomology groups\footnote{Here we always consider singular (co)homology with integer coefficients unless otherwise specified.}
\begin{equation}\label{eqn:K}
K^1(\partial \mathcal{M}_{10})=\bigoplus_i H^{2i+1}(\partial \mathcal{M}_{10})\, .
\end{equation}

In analogy with QFT, we can assign a Hilbert space $\mathcal{H}(\partial \mathcal{M}_{10})$ to the boundary of $\mathcal{M}_{10}$ at infinity. Fixing a boundary condition at infinity for the fluxes corresponds to picking a specific state in the Hilbert space.
The non-commutativity of fluxes implies that we cannot fix the boundary conditions, that is, expectation values for the fluxes at infinity, for all the fluxes at the same time.  

As a result, the best option would be to pick a maximal isotropic subgroup $L$ of ${\Tor K^1(\partial \mathcal{M}_{10})}$ such that for all the $\sigma\in L$ the corresponding operators $\phi_\sigma$ commute. That is, we are selecting a maximal set of mutually local discrete fluxes and decompose the Hilbert space in terms of their eigenvalues.

Here, we can take $\mathcal{W}_6$ to be a cone over a manifold $\mathcal{W}_{5}$ such that
\begin{equation}\label{boundary}
    \partial\mathcal{M}_{10}=\mathcal{M}_4\times \mathcal{W}_5\, .
\end{equation} 
In order to produce a Hilbert space, we want to perform canonical quantisation of type IIB on the Euclidean background $\mathcal{M}_{10}$, where asymptotically $\mathcal{M}_{10}=\mathbb{R}\times\partial \mathcal{M}_{10}$, by choosing time to be the radial direction, i.e. the direction normal to the boundary. This means the boundary $\partial \mathcal{M}_{10}$ at infinity would be the slice with the radius of the cone going to infinity.

The non-commutativity of fluxes due to torsion implies that the Hilbert space is graded by the free abelian group of fluxes modulo torsion. More precisely, as derived in \cite{Freed:2006ya, Freed:2006yc} where we refer the reader for more details, the non-commutativity of fluxes gives rise to the grading of the Hilbert space $\mathcal{H}_\alpha$ by the cosets
\begin{equation}
    \alpha\in\frac{K^1(\partial \mathcal{M}_{10})}{\Tor K^1(\partial \mathcal{M}_{10})}\, .
\end{equation} 

The partition functions $\bra{Z}$ form a basis of the vector space dual to the Hilbert space $\mathcal{H}_\alpha^*(\partial \mathcal{M}_{10})$, such that for any $\ket{\phi}\in\mathcal{H}_\alpha(\partial \mathcal{M}_{10})$, we have $\braket{Z\mid\phi}\in \mathbb{C}$. Hence, by just fixing $L$, to say $L_0$, we have a set of partition functions, i.e. a partition vector. In order to fix the partition function, we must pick a state $\ket{0,L_0}\in \mathcal{H}_\alpha(\partial \mathcal{M}_{10})$ in the Hilbert space to have a fixed eigenvalue, say a unit 
\begin{equation}
    \phi_{\sigma'} \ket{0,L_0}=\ket{0,L_0}    
\end{equation}
under the action of all flux operators for all $\sigma'\in L_0$. In other words, we fix the state with ``zero'' flux, which corresponds to fixing a boundary condition at infinity for the fluxes. Now, the action of $ \phi_{\sigma}$, with $\sigma\in{\Tor K^1(\partial \mathcal{M}_{10})}/{L}$ a representative of a different maximal isotropic subspace $L$, gives
\begin{equation}
    \phi_{\sigma} \ket{0,L_0}=\ket{\sigma,L_0}    
\end{equation}
and the set of all representatives $\sigma$ form a basis $\{\ket{\sigma,L_0}: \sigma\in{\Tor K^1(\partial \mathcal{M}_{10})}/{L}\}$ of the Hilbert space. That is, the Hilbert space has a basis 
given by 
\begin{equation}
    {\Tor K^1(\partial \mathcal{M}_{10})}/{L}\, .
\end{equation}
Note that the fluxes $\phi_{\sigma'}$ with $\sigma'\in L_0$ are precisely the diagonal ones in this basis. As such, by equation (\ref{noncommute})
\begin{equation}
    \phi_{\sigma'} \ket{\sigma,L_0}=e^{2\pi i L(\sigma,\sigma')}\ket{\sigma,L_0}\, .    
\end{equation}

Not all theories have a maximally isotropic subgroup $L$ that is invariant under large diﬀeomorphisms of the boundary $\partial \mathcal{M}_{10}$. This means the partition vector
$\bra{Z}=\sum_i z_i\ket{\sigma_i,L_0}^*$ defined in terms of the dual basis vectors $\ket{\sigma_i,L_0}^*$ of the Hilbert space, and consequently the partition function which depend on the choice of $L$,
are also not invariant under such transformations.

\subsection{Defect groups from BPS quivers}\label{defectcomp}

Let us now discuss the higher form symmetries  of $\mathcal{T}_{\mathcal{W}_6}$ theories and find the corresponding defect groups by summarising the results of \cite{4dpaper} for completeness. 

As stated in the previous section, the ﬂuxes in IIB string theory are classiﬁed by the cosets of the K-theory group $K^1(\partial \mathcal{M}_{10})$. In our case, the only operators $\phi_\sigma$ with non-trivial commutation relations have $\sigma\in \Tor H^5(\partial \mathcal{M}_{10})\subset K^1(\partial \mathcal{M}_{10})$.  
By the Künneth formula (\ref{eqn:K}) this reduces to
\begin{equation}
    \Tor H^5(\partial \mathcal{M}_{10})=
    H^{2}(\mathcal{M}_4) \otimes \Tor H^3(\mathcal{W}_5),
\end{equation}
since $H^3(\mathcal{W}_5)$ is the only non-vanishing torsional cohomology group of $\mathcal{W}_5$ \cite{Milnor}. Hence, in order to find the defect group associated to the 1-form symmetry of $\mathcal{T}_{\mathcal{W}_6}$, we need to find the homology groups of $\mathcal{W}_5$. 

We are interested in geometries where $\mathcal{W}_6$ is an algebraic hypersurface defined by a polynomial $P$ \begin{equation}\label{eq:m6}
\mathcal{W}_6= \{\mathbf{x} \in \mathbb{C}^4 \mid P(\mathbf{x})=0\}\df \{P=0\}
\end{equation}
with an isolated singularity at the origin. Define $\mathcal{Y}_5=S^7\cap\{P=\epsilon\}$ to be a 5-dimensional manifold that is homotopy equivalent to the boundary at infinity $ \mathcal{W}_5$, where $\epsilon$ is small enough to make $\mathcal{Y}_5$ smooth and $S^7$ is a small 7-dimensional sphere centered at origin in $\mathbb{C}^4$. Given that the interior $\mathcal{Y}_6$ of $\mathcal{Y}_5=\partial \mathcal{Y}_6$ is homotopy equivalent to a bouquet of 3-spheres \cite{Milnor}, we have a long exact sequence of relative homology 
\begin{equation}
\ldots \to H_3(\mathcal{Y}_6) \xrightarrow {Q} H_3(\mathcal{Y}_6,\mathcal{Y}_5) \to H_2(\mathcal{Y}_5) \to 0\, ,
\end{equation}
where the map $Q$ is isomorphic to the intersection form of 3-cycles in $\mathcal{Y}_6$. 
Therefore, the defect group $\mathbb{D}$ is \cite{Caorsi:2017bnp}
\begin{equation}\label{eq:D}
    \mathbb{D}=H^3(\mathcal{Y}_5)=H_2(\mathcal{Y}_5) = \frac{H_3(\mathcal{Y}_6,\mathcal{Y}_5)}{Q(H_3(\mathcal{Y}_6))}= \text{coker} (Q)\, ,
\end{equation} 
where the second equality is implied by Poincaré duality. That is, the defect group may be understood as the D3 branes wrapping 3-cycles in $\mathcal{Y}_6$ which give rise to lines in $\mathcal{T}_{\mathcal{W}_6}$, modulo the D3 branes wrapping compact 3-cycles in $\mathcal{Y}_6$  which give rise to dynamical lines in $\mathcal{T}_{\mathcal{W}_6}$.

The intersection form $Q$, known as the Dirac pairing in the language of BPS quivers, can be easily read from the BPS quivers as follows. In type IIB string theory the set of nodes $\{i\}$ in the BPS quiver form a basis of $3$-cycles in $\mathcal{Y}_6$ on which the D3 branes wrap, and so the number of arrows pointing from each node $i$ to $j$, denoted $q_{ij}=-q_{ji}\, $, give the intersection number of the $3$-cycle at node $i$ with the one at $j$  \cite{Aspinwall:2004jr}. That is, the matrix $\{q_{i,j}\}$ is a representation of the map $Q$.

Recall that if $Q$ is a matrix with entries in $\mathbb{Z}$, then there are invertible matrices $A$ and $B$ over $\mathbb{Z}$ such that $Q=AQ_{SNF}B$, where $Q_{SNF}=\text{diag}\{a_1,a_2,a_3,\hdots,a_n,0,\hdots,0\}$, such that $a_i$ are integers and $a_i$ divides $a_{i+1}$ for each $i < n$, is called the Smith normal form of $Q$. As this is only a change of basis, the matrices $Q$ and $Q_{SNF}$ have the same cokernel given by 
\begin{equation}
    \text{coker} (Q)=\text{coker} (Q_{SNF})=\mathbb{Z}^f\oplus \mathbb{Z}^n/(a_1\mathbb{Z}\oplus  a_2\mathbb{Z}\oplus\hdots\oplus a_n\mathbb{Z})\, ,
\end{equation}
with $f$ the number of zero diagonal elements of $Q_{SNF}$. From the definition of $\mathrm{coker}(Q)$, the $\mathbb{Z}^f$ factor corresponds to vectors which lie in $\mathrm{ker}(Q)$. As a flavour charge has null Dirac pairing with all other charges in the charge lattice, $\mathrm{ker}(Q)$ is exactly the space of flavour charges of the four dimensional theory. It is then clear that $f$ is the rank of the flavour group while $n/2$ is rank of the engineered theory.

Moreover, it is important to note that in our calculations we expect the $a_i$ to come in pairs, i.e. $a_1=a_2,\; a_3=a_4,\; \hdots,\; a_{n-1}=a_n$. Thus, renaming the labels $i$, we must have
\begin{equation}
    \Tor \big(\text{coker} (Q)\big)=
    \bigoplus_{i=1}^{n/2}\; (\mathbb{Z}_{a_i} \oplus \mathbb{Z}_{a_i})\, .
\end{equation}
This can be understood in terms of the linking pairing (\ref{linking}) determining the commutation relations (\ref{noncommute}). We do not calculate the linking pairing here, however we can predict the above from the field theory perspective. Specifically, by analogy to 4d Maxwell's or Yang-Mills theory studied in \cite{Aharony:2013hda}, one factor corresponds to the theory with electric 1-form symmetry with higher group $\oplus_i \mathbb{Z}_{a_i}$ and similarly the other to the magnetic theory.\footnote{Note that, it is possible to have chosen a different maximal isotropic subgroup $L$, such that for example the higher group is a diagonal subgroup of $\mathbb{D}$, in which case we would instead obtain a dyonic symmetry.} It should be noted that, there is not always such a splitting of the torsional defect group into a magnetic and electric part, see for example the 2-form symmetries in 6d $\mathcal{N}=(1,0)$ theories \cite{DelZotto:2015isa} or $\mathcal{N}=(2,0)$ theories \cite{Garcia-Etxebarria:2019cnb}.

\subsection{Defect group from Orlik's conjecture}\label{OrlikC}
Apart from reading the defect groups from the $\text{coker}(Q)$ of the BPS quivers, the defects groups (\ref{eq:D}) of $\mathcal{T}_{\mathcal{W}_6}$ theories may also be determined in the exact form using Orlik's conjecture \cite{OM} and its proof \cite{Boyer}, as discussed in \cite{4dpaper}. Explicitly, let the polynomial $P=P(x,y,z,w)$ defining  $\mathcal{W}_6=\{P=0\}$ be a weighted-homogeneous polynomial of degree $d$ and weights $(w_1, w_2,w_3,w_4)$, i.e. $P(\lambda^{w_1}x,\lambda^{w_2}y,\lambda^{w_3}z,\lambda^{w_4}w)=\lambda^d P(x,y,z,w)$. Then, the torsional part of the defect group is
\begin{equation}
  \label{eq:BGS}
  \Tor \mathbb{D}=\Tor H_2(Y_5)=\sum_{i=1}^4\mathbb{Z}_{r_i}^{2g_i}\, ,
\end{equation}
where $r_i=\text{gcd}(w_1,..,\hat{w}_i,...,w_4)$, the exponents $2g_i$ are by
\begin{equation}
    2g_i=-1+\sum_{j\neq i}\frac{\text{gcd}(d,w_j)}{w_j}-
d\sum_{j< k\, ,\, j,k\neq i}\frac{\text{gcd}(w_j,w_k)}{w_jw_k}+\frac{d^2r_i w_i}{w_1 w_2 w_3 w_4}\, ,
\end{equation}
and the notation $\;\hat{}\;$ means that the corresponding
element in the list should be omitted.

Furthermore, the dimension of the matrix $Q$ defining the BPS quiver is given by the Milnor number $\mu=\mu(P)$ of the hypersurface singularity \cite{OM}
\begin{equation}\label{MN}
    \mu=\Big(\frac{d}{w_1}-1\Big)\Big(\frac{d}{w_2}-1\Big)\Big(\frac{d}{w_3}-1\Big)\Big(\frac{d}{w_4}-1\Big)\, .
\end{equation}

\section{Defect groups of $D_p(G)$ theories}
\subsection{Summary of $D_p(G)$ theories}
When we geometrically engineer a theory, it is typical to take the singular hypersurface to be embedded in $\mathbb{C}^4$. However, we can also consider the possibility that one or more of the directions of the hypersurface is represented by a $\mathbb{C}^\times$ variable. An example of such a geometry is given by
\begin{gather}\label{eqn:pure}
    \mathcal{W}_{\mathrm{pure}} =\{\mathbf{x}\in \mathbb{C}^4: U^2+X^2+Y^2+e^Z+e^{-Z}=u\},
\end{gather}
where $\mathbf{x}=(U,X,Y,Z)$ and $u$ is a parameter that descends to a Coulomb branch parameter in the four dimensional theory. It is known that this geometry engineers pure $\text{SU}(2)$ super Yang-Mills (SYM) and the resulting BPS quiver is the $A(1,1)$ affine Dynkin quiver \cite{Cecotti:1992rm, Cecotti:2010fi}.

One can consider generalisations of \eqref{eqn:pure} to encounter more theories with affine quiver components. In particular, consider the geometry given by
\begin{gather}\label{eqn:affgeo}
        \mathcal{W} =\{\mathbf{x}\in \mathbb{C}^4: U^2+W_G(X,Y)+e^{pZ}+e^{-Z}=0\},
\end{gather}
where $W_G$ is the $G$-type Du Val singularity\footnote{More precisely, it is the the versal deformation of the minimal  $G$-type Du Val singularity. Up to analytic isomorphism, it can be found by removing the squared term of  each  Du Val singularity.} and $p\in \mathbb{N}$. As the $(X,Y)$ and $Z$ terms aren't mixed, the resulting four dimensional theory is factored in the same sense as $(G, G')$ theories. The BPS quivers for the theories compactified on this geometry are simply given by $A(p,1)\boxtimes G$.

By studying the light subcategory of these theories, it was shown in \cite{dpg1} that there exists a corner of parameter space where they simplify to SYM with gauge group $G$ coupled to a superconformal system  which we call the $D_p(G)$ theory. Decoupling the SYM factor, by taking the SYM coupling to zero, leaves us with the $D_p(G)$ theory with an enhanced flavour group containing $G$. In particular, the rank of the resulting theory's flavour group is given by
\begin{gather}
	\mathrm{rank}\; F = \mathrm{rank}\; G +\sum_{d \in I_n^p} \varphi(d)= f(p; G),
\end{gather}
where $\phi$ is the Euler totient function and $I_n^p$ is a subset of divisors of $p$ and $h^\vee(G)$. Furthermore, the effect of decoupling the $G$-SYM is manifest on both the BPS quiver and the engineering geometry. The geometry which engineers the $D_p(G)$ theory is obtained from \eqref{eqn:affgeo} by simply dropping the $e^{-Z}$ term while the BPS quiver reduces to $A(p,0) \boxtimes G$ \cite{dpg2}, the intersection matrix of which can be usefully written as
\begin{gather}\label{eqn:intform}
	B = (1-P)\otimes S_G + (1-P)^T \otimes S_G^T,
\end{gather}
where $P$ is the cyclic permutation matrix acting on $p$ elements and $S_G$ is the Stokes matrix of $G$ \cite{Cecotti:1992rm}.
\subsection{Defect group and Maruyoshi-Song flows}

\begin{table}[]
\centering
\begin{tabular}{|c|cl|}
\hline
\rowcolor[HTML]{DAE8FC} 
\textbf{AD theory $\mathcal{T}$} & \multicolumn{2}{c|}{\cellcolor[HTML]{DAE8FC}\textbf{Torsional part of defect group $ \mathrm{Tor}\;\mathbb{D}_\mathcal{T}$}}                                                                                                                                          \\ \hline
$(A_n, A_m)$                     & \multicolumn{2}{c|}{$0$}                                                                                         \\ \hline
$(A_{n-1}, D_{m+1})$             & \begin{tabular}[c]{@{}c@{}}$\mathbb{Z}_2^{\mathrm{gcd}(n,m)-1}$\\ $\mathbb{Z}_2^{\mathrm{gcd}(n,m)-2}$\\ $0$\end{tabular} & \multicolumn{1}{l|}{\begin{tabular}[c]{@{}l@{}}if $2\nmid n$ \\ if $2\mid n,m$ and $\mathrm{gcd}(n,2m)\mid m$ \\ otherwise\end{tabular}}  \\ \hline
$(A_{n-1}, E_6)$                 & \begin{tabular}[c]{@{}c@{}}$0$\\ $\mathbb{Z}_2^2$\\ $\mathbb{Z}_3^2$\\ $0$\end{tabular}                                   & \multicolumn{1}{l|}{\begin{tabular}[c]{@{}l@{}}if $12 \mid n$\\ if $6 \mid n$\\ if $4 \mid n$\\ otherwise\end{tabular}}                   \\ \hline
$(A_{n-1}, E_7)$                 & \begin{tabular}[c]{@{}c@{}}$0$\\ $\mathbb{Z}_2^6$\\ $\mathbb{Z}_3^2$\\ $0$\end{tabular}                                   & \multicolumn{1}{l|}{\begin{tabular}[c]{@{}l@{}}if $18 \mid n$\\ if $9 \mid n$\\ if $6 \mid n$\\ otherwise\end{tabular}}                   \\ \hline
$(A_{n-1}, E_8)$                 & \begin{tabular}[c]{@{}c@{}}$0$\\ $\mathbb{Z}_2^8$\\ $\mathbb{Z}_3^4$\\ $\mathbb{Z}_5^2$\\ $0$ \end{tabular}      & \multicolumn{1}{l|}{\begin{tabular}[c]{@{}l@{}}if $30 \mid n$\\ if $15 \mid n$\\ if $10 \mid n$\\ if $6 \mid n$\\ otherwise\end{tabular}} \\ \hline
\end{tabular}
\caption{The torsional parts of the defects groups for theories of type $(\mathfrak{g},\mathfrak{g}')$. For theories where the cases overlap, the highest written condition takes priority.}\label{tbl:gg'defects}
\end{table}

If we use an engineering geometry of the form \eqref{eqn:affgeo}, we cannot use the methods of \cite{Boyer} to calculate the defect group as the geometry is not manifestly quasi-homogeneous. One can instead use the substitution $t=e^Z$ to obtain an equation that is quasi-homogeneous. These geometries coincide with those of $(A_{p-1},G)$ with some caveats. The most important of which is that the change of variable alters not only the geometry, but also the holomorphic top-form $\Omega^{\mathcal{T}}$ of the theory $\mathcal{T}$. In fact, the substitution is such that
\begin{gather}
    \Omega^{D_p(G)}=\frac{\Omega^{(A_{p-1},G)}}{t}.
\end{gather}
As we must have that $[\Omega^{\mathcal{T}}]=1$, this explicitly changes the scaling dimensions of the geometry. This shows that these two classes of theories are indeed different.

Motivated by this, one can then postulate that the defect groups of $D_p(G)$ and $(A_{p-1}, G)$ theories have the same torsional part, and different flavour factors. However, as Orlik's algorithm \cite{OM} is typically formulated for surfaces in $\mathbb{C}^n$ we should be cautious\footnote{If one assumes that the algorithm holds exactly in this case, one would use it to find the flavour ranks of the theories and conclude that they coincide. This is not the case!}. We therefore test this hypothesis by instead consulting the BPS quivers of the theories. 

Using \eqref{eqn:intform} as the intersection form we can use the Smith normal form of $B$ to read off the cokernel as described in section \ref{defectcomp}. Doing so for $3\leq p \leq 30$ and $3\leq \mathrm{rank}\; G \leq 30$ we notice that the torsional part of the defect group is that of the $(A_{p-1}, G)$ theory, as expected. Indeed, we have
\begin{gather}
	\mathbb{D}_{D_p(G)}= \mathbb{Z}^{f(p; G)}\oplus \mathrm{Tor}\; 
	\mathbb{D}_{(A_{p-1},G)}.
\end{gather}
This is a realisation of the Maruyoshi-Song (MS) flow from $D_p(G)$ to $(A_{p-1}, G)$ \cite{Giacomelli:2017ckh, Giacomelli:2020ryy}. Furthermore, we already know the defect groups for the $(A_{p-1}, G)$ theories, so we can simply read them from Table \ref{tbl:gg'defects}.

Evidence of this MS flow can been seen at the level of BPS quivers. The $D_p(G)$ theory possesses a BPS quiver of size $p\cdot\mathrm{rank}\;G$ given by $A(p,0)\boxtimes G$. Deforming the theory by an MS term and triggering a flow using the principal nilpotent VEV of $\mathfrak{g}$ will reduce the size of the quiver in the IR to $|Q_{IR}|=(p-1)\, \mathrm{rank}\; G$ \cite{bpsrg}. Furthermore, the IR quiver will have nullity given by
\begin{gather}\label{eqn:flav}
	\mathrm{dim}(\mathrm{ker}\; B_{IR})=\mathrm{rank}\; F- \mathrm{rank}\; G=\sum_{d \in I_n^p} \varphi(d).
\end{gather} 
This is a consequence of triggering the flow with the principal nilpotent VEV of $\mathfrak{g}$ instead of the full flavour symmetry  $\mathfrak{f}$ of $D_p(G)$. Detaching one full $G$ factor in the BPS quiver will break the $A(p,0)$ to $A_{p-1}$ and leave us with $A_{p-1}\boxtimes G$, a possible BPS quiver for $(A_{p-1}, G)$, which has the correct size to be a candidate for the IR theory. Let us check the rank of the flavour group for these theories.

\begin{table}[]
\centering
\begin{tabular}{|c|c|c|}
\hline
\rowcolor[HTML]{DAE8FC} 
\textbf{Group} & \textbf{Coxeter number $h(G)$} & \textbf{Characteristic polynomial of $\Phi_G$} \\ \hline
$A_n$          & $n+1$           & $t^n+t^{n-1}+\ldots+t+1$                       \\
$D_n$          & $2n-2$          & $(t+1)(t^{n-1}+1)$                             \\
$E_6$          & $12$            & $(t^2+t+1)(t^4-t^2+1)$                         \\
$E_7$          & $18$            & $(t+1)(t^6-t^3+1)$                             \\
$E_8$          & $30$            & $t^8+t^7-t^5-t^4-t^3+t+1$                      \\ \hline
$A(p,1)$	   & $-$			 & $(t-1)(t^p-1)$
\\ \hline
\end{tabular}
\caption{The Coxeter numbers of simply-laced Dynkin groups and the characteristic polynomials of their Coxeter elements \cite{coxeterchar}. Additionally, we list the characteristic polynomial for the Coxeter element of $A(p,1)$ \cite{dpg2}.}\label{tbl:charpol}
\end{table}

We can infer the flavour rank of the theory by inspecting the eigenvalues of the 2d monodromy $H= (S^{-1})^T S$ formed from the Stokes matrix $S$ of the quiver \cite{Cecotti:1992rm}. For the $(G, G')$ theories, we can write this as
\begin{gather}
	H= \Phi_{G} \otimes \Phi_{G'},
\end{gather}
where $\Phi_G$ is the Coxeter element of the group $G$. The flavour rank of the theory is given by the number of eigenvalues of $H$ that are equal to one. We know that the characteristic polynomial is given by
\begin{gather}\label{eqn:diophantine}
	\mathrm{det}\;\Big(H-t\; \mathrm{id}_{r(G) \times r(G')}\Big)\,=\,\prod_{i=1}^{r(G)} \prod_{j=1}^{r(G')}\Big(t-\lambda^{G}_i\lambda^{G'}_j\Big)\, ,
\end{gather}
where $\lambda^{G}_i$ are the eigenvalues of $\Phi_G$. The problem is now reduced to finding appropriate products of eigenvalues for $G= A_{p-1}$. We list the characteristic polynomials of the $ADE$ group's Coxeter elements in Table \ref{tbl:charpol} from which we can infer the eigenvalues. 

Condition \eqref{eqn:diophantine} imposes a linear Diophantine equation for each choice of $G'$ from which one can, in principle, establish the results. An easier method is to instead notice that the characteristic polynomial for $A(p,1)$ is simply that of $A_{p-1}$ with two additional roots at $t=1$. However, no other characteristic polynomial listed in Table \ref{tbl:charpol} has a root at $t=1$, so no product with the additional roots will contribute to the flavour rank. We therefore have
\begin{gather}
	F((A_{p-1}, G))=F(A(p,1)\boxtimes G),
\end{gather} 
where $F(\cdot)$ denotes the flavour rank of the theory. Noting that the additional flavour symmetries in the $D_p(G)$ theory (that is, the terms on the RHS of equation \eqref{eqn:flav}) are exactly those flavour symmetries present in the $A(p,1)\boxtimes G$ theory, we have that the $(A_{p-1}, G)$ theory possesses exactly the desired flavour rank. Piecing these facts together, we see that this signals a possible MS flow 
\begin{gather}
	D_p(G) \longrightarrow (A_{p-1}, G)
\end{gather}
for any $p>1$ and simply-laced $G$. This is in agreement with the MS flow found in \cite{Giacomelli:2017ckh, Giacomelli:2020ryy}.
 
\section{Defect groups of $D_p^b(G)$ theories}
In the previous section, we considered theories which are engineered by a surface including a Du Val singularity and an affine ($\mathbb{C}^\times$) part. We can further generalise these theories by replacing the singularity by a compound Du Val singularity \cite{reid1983}. We start with the $\mathbb{C}^4$ hypersurface defined by the vanishing locus of
\begin{gather}
    W^{\mathrm{comp.}}_G(u,x,y,z)= u^2+W_G(x, y)+zg(u,x,y,z),
\end{gather}
where $W_G(x, y)$ is the equation of the $G$-type Du Val singularity and $g$ is an arbitrary polynomial in four variables. However, it is known that not all compound Du Val singularities are isolated \cite{dais2001resolving} so requiring this be the case greatly restricts the geometry. In fact, it can be shown that the only possible isolated geometries in this case are the ones listed in Table \ref{tbl:cdv} \cite{Wang:2015mra}.
\begin{table}[h!]
\centering
\begin{tabular}{|c|c|c|}
\hline
\rowcolor[HTML]{DAE8FC} 
\textbf{Group} & \textbf{$b$} & \textbf{Singular hypersurface $\mathcal{W}\subset \mathbb{C}^4$} \\ \hline
$A_n$          & $n+1$        & $u^2+x^2+y^{n+1}+z^k=0$                                                      \\
               & $n$          & $u^2+x^2+y^{n+1}+yz^k=0$                                                         \\ \hline
$D_n$          & $2n-2$       & $u^2+x^{n-1}+xy^2+z^k=0$                                                     \\
               & $n$          & $u^2+x^{n-1}+xy^2+yz^k=0$                                                    \\ \hline
$E_6$          & 12           & $u^2+x^3+y^4+z^k=0$                                                          \\
               & 9            & $u^2+x^3+y^4+yz^k=0$                                                         \\
               & 8            & $u^2+x^3+y^4+xz^k=0$                                                         \\ \hline
$E_7$          & 18           & $u^2+x^3+xy^3+z^k=0$                                                         \\
               & 14           & $u^2+x^3+xy^3+yz^k=0$                                                        \\ \hline
$E_8$          & 30           & $u^2+x^3+y^5+z^k=0$                                                          \\
               & 24           & $u^2+x^3+y^5+yz^k=0$                                                         \\
               & 20           & $u^2+x^3+y^5+xz^k=0$                                                         \\ \hline
\end{tabular}
\caption{The singular hypersurfaces which exhibit {\it isolated} compound Du Val singularities. The theories engineered from these geometries are called the $G^{(b)}[k]$ theories and taking $z\mapsto e^Z$ gives the $D_k^b(G)$ theories.}\label{tbl:cdv}
\end{table}

These geometries give rise to the $G^{(b)}[k]$ theories of \cite{Wang:2015mra} and taking $z=e^Z$ gives the $D_k^b(G)$ theory. The parameter $b$ takes on physical meaning in the $G^{(b)}[k]$ theory where the holomorphic part of the Higgs field of the associated Hitchin system has asymptotic form
\begin{gather}
    \Phi_z\sim \frac{T}{z^{2+k/b}},
\end{gather}
where $T$ is some regular semi-simple element of $G$.

Our goal is to compute the defect groups of both the $D_p^b(G)$ and the $G^{(b)}[k]$. Since we know that there is a Maruyoshi-Song flow for any $D_k^b(G)$ theory taking it to $G^{(b)}[k]$ triggered by the principal nilpotent VEV of $G$ \cite{Giacomelli:2017ckh}\footnote{Indeed, the flow mentioned in the previous section is of this form.} the problem is greatly simplified. From this we can find the defect groups of $D_k^b(G)$ theories from those of the $G^{(b)}[k]$ which in turn can be found easily via Orlik's theorem. We then use the polar curve method of Gabrielov \cite{gabrielov} to find the BPS quivers of some special cases and use these to verify our calculations.

Furthermore, note that some of the theories discussed above have a Lagrangian description \cite{Agarwal:2017roi, Benvenuti:2017bpg, Carta:2020plx}. Thus, it is also possible to find the defect groups by looking at their gauge groups and matter contents. In fact, these cases including the $D_n^{(n)}[k]$ theories are discussed in \cite{4dpaper}.

\subsection{Orlik's algorithmic approach}

As discussed in section \ref{OrlikC}, for theories obtained by putting type IIB on a hypersurface defined by a weighted-homogeneous polynomial, the defect group is given by equation (\ref{eq:BGS}) which is copied here
\begin{equation}
  \Tor H_2(Y_5)=\sum_{i=1}^4\mathbb{Z}_{r_i}^{2g_i}\, ,
\end{equation}
where $r_i=\text{gcd}(w_1,..,\hat{w}_i,...,w_4)$, the exponents $2g_i$ are given by 
\begin{equation}
    2g_i=-1+\sum_{j\neq i}\frac{\text{gcd}(d,w_j)}{w_j}-
d\sum_{j< k\, ,\, j,k\neq i}\frac{\text{gcd}(w_j,w_k)}{w_jw_k}+\frac{d^2r_i w_i}{w_1 w_2 w_3 w_4}\, ,
\end{equation}
and the notation hat means that the corresponding
element in the list should be omitted.

Therefore, given $J^{(b)}[k]$ theories are defined by the hypersurfaces presented in Table \ref{tbl:cdv}, we apply this formula to obtain the corresponding defect groups and list them in Table \ref{tbl:Gb(k)defects}.

To illustrate how to apply the above formula, let us look at the example of $D^{(n)}_n[k]$ case. Here the corresponding polynomial is 
$$W=u^2+x^{n-1}+xy^2+yz^k\,.$$

To find the weights $w$ and the degree $d$ of this polynomial\footnote{For a full classification see \cite{2003math......3302Y}.}, we recall that they must satisfy  $W(\lambda^{w_1}u,\lambda^{w_2}x,\lambda^{w_3}y,\lambda^{w_4}z)=\lambda^{d} W(u,x,y,z)$. Hence, the weights must have the form $$w=\Big(w_1,\frac{2w_1}{n-1},\frac{2w_1(n-2)}{2(n-1)},\frac{2nw_1}{2(n-1)k}\Big)\, , \quad\text{where} \quad w_1=\text{lcm}\Big(1,\frac{n-1}{2},\frac{n-1}{n-2},\frac{k(n-1)}{n}\Big)=\frac{k(n-1)}{\gcd(k,n)}\, .$$ That is, we have
\begin{equation}
    w=\big(k(n-1),2k,k(n-2),n\big)/\gcd(k,n)\, \quad\text{and}\quad d=2k(n-1)/\gcd(k,n)\, .
\end{equation}
\begin{table}[]
\centering
\begin{tabular}{|c|cl|}
\hline
\rowcolor[HTML]{DAE8FC} 
\textbf{Theory $\mathcal{T}$} & \multicolumn{2}{c|}{\cellcolor[HTML]{DAE8FC}
\textbf{Torsional part of defect group $\mathrm{Tor}\;\mathbb{D}_\mathcal{T}$}} 
\\ [0.1cm]
\hline  
$A_{n-1}^{(n-1)}[k]$                   
& \begin{tabular}[c]{@{}c@{}} $0$\end{tabular} 
& \multicolumn{1}{l|}{\begin{tabular}[c]{@{}l@{}}  \end{tabular}} \\ [0.1cm]
\hline

$D_{n}^{(n)}[k]$             
& \begin{tabular}[c]{@{}c@{}}$ \mathbb{Z}_2^{\mathrm{gcd}(2k,n)-2}$\\ $0$\end{tabular} 
& \multicolumn{1}{l|}{\begin{tabular}[c]{@{}l@{}}if  $2\mid\frac{n}{\gcd(k,n)}$ \\ otherwise\end{tabular}} \\ [0.1cm]
\hline
$E_6^{(9)}[k]$                 
& \begin{tabular}[c]{@{}c@{}}  $\mathbb{Z}_3^2$\\  $0$\end{tabular}                          
& \multicolumn{1}{l|}{\begin{tabular}[c]{@{}l@{}}if $9 \nmid k$ and $3\mid k$\\ otherwise\end{tabular}}               
\\ [0.1cm]
\hline
$E_6^{(8)}[k]$                 
& \begin{tabular}[h!]{@{}c@{}} $\mathbb{Z}_2^2$\\ $0$\end{tabular}                   
& \multicolumn{1}{l|}{\begin{tabular}[c]{@{}l@{}}if $8 \nmid k$ and $4 \mid k $\\  otherwise\end{tabular}}              
\\ [0.1cm]
\hline
$E_7^{(14)}[k]$                 
& \begin{tabular}[c]{@{}c@{}} $\mathbb{Z}_2^6$\\ $0$ \end{tabular}      
& \multicolumn{1}{l|}{\begin{tabular}[c]{@{}l@{}}if $2 \nmid k$ and $7 \mid k$\\ otherwise\end{tabular}} 
\\ [0.1cm]
\hline

$E_8^{(24)}[k]$                 
& \begin{tabular}[c]{@{}c@{}} 
$\mathbb{Z}_2^8$\\$\mathbb{Z}_2^4$\\ $\mathbb{Z}_2^2$\\ [0.1cm] $\mathbb{Z}_3^4$\\ [0.1cm] $0$ \end{tabular}

& \multicolumn{1}{l|}{\begin{tabular}[c]{@{}l@{}}  if $24 \nmid k$ and $12 \mid 4$\\  if $6 \mid k$\\  if $3 \mid k$\\ [0.1cm] if $8 \mid k $\\
otherwise\end{tabular}} 
\\ [0.1cm]
\hline
$E_8^{(20)}[k]$                 
& \begin{tabular}[c]{@{}c@{}}
$\mathbb{Z}_2^8$\\ [0.1cm]
$\mathbb{Z}_2^4$\\ [0.1cm] $\mathbb{Z}_5^2$\\ $0$ \end{tabular}
& \multicolumn{1}{l|}{\begin{tabular}[c]{@{}l@{}}  if $20 \nmid k$ and $10 \mid k$\\ [0.1cm] if $5 \mid k$\\ [0.1cm] if $4 \mid k$\\ [0.1cm] otherwise\end{tabular}} 
\\ [0.1cm]
\hline

\end{tabular}
\caption{The torsional parts of the defects groups for $J^{(b)}[k]$ theories. For theories where the cases overlap, the highest written condition takes priority.}\label{tbl:Gb(k)defects}
\end{table}
Now to find the defect group it remains to find $r_i$ and $g_i$. For example, $r_1=\gcd(w_2,w_3,w_4)=\gcd(2k,n)/\gcd(k,n)$ and 
\begin{equation}
\begin{split}
    2g_1=&-1+\frac{\gcd(d,w_2)}{w_2}+\frac{\gcd(d,w_3)}{w_3}+\frac{\gcd(d,w_4)}{w_4}-d\frac{\gcd(w_2,w_3)}{w_2w_3}-\\&-d\frac{\gcd(w_2,w_4)}{w_2w_4}-d\frac{\gcd(w_3,w_4)}{w_3w_4}+\frac{d^2r_1}{w_2w_3w_4}\\
    =&\gcd(2k,n)-2
\end{split}
\end{equation}
for even $n$. Similarly, we find that all the other $\mathbb{Z}_{r_i}^{2g_i}$ for $i\neq1$ vanish. Therefore, we have

\begin{equation}
\Tor H_2(Y_5)=
    \begin{cases}
    \mathbb{Z}_2^{\mathrm{gcd}(2k,n)-2}       & \quad \text{if } 2\mid\frac{n}{\gcd(k,n)}\\
    0  & \quad \text{Otherwise. }
  \end{cases}
\end{equation}

\subsection{BPS quivers for $A_n^{(n)}[k]$ theories}
To cross check some of our results we can look at the BPS quivers of the theories. To do so, we need to construct them using the polar curve method \cite{gabrielov, ebeling2006monodromy}. This is reviewed briefly in appendix \ref{ap:polar} to which we direct the unfamiliar reader before proceeding with the rest of this section. Heuristically, the method breaks down the problem into a reduced singularity and a set of integers calculated from the singularity. The integers encode how to build the full Coxeter-Dynkin diagram of the singularity which we illustrate in the case of $A_{n}^{(n)}[k]$.

The engineering geometry for the $A_{n-1}^{(n-1)}[k]$ theory is given by the vanishing locus of 
\begin{gather}
    \xi=u^2+x^2+y^n+y z^k.
\end{gather}
As the first two terms are just squares, we can disregard these terms and consider the stably equivalent singularity $f=y^n+y z^k$ so that $f|_{z=0}=y^n$, an $A_{n-1}$ singularity, and hence $\mu(f|_{z=0})=n-1$. The critical points of $F_\varepsilon$ are given by
\begin{gather}
    y^*_m=\big[e^{(2m+1)\pi i}\varepsilon^{k}n^{-1}\big]^{1/(n-1)}
    \Rightarrow F^*_\varepsilon=\bigg(1-\frac{1}{n}\bigg)\big(e^{(2m+1)\pi i}\varepsilon^{kn}n^{-1}\big)^{1/(n-1)},
\end{gather}
where $m=0,\ldots, n-2$. Therefore, the critical values of $F_\varepsilon$ form a regular $(n-1)$-gon centred on the origin. We label these roots in order of decreasing argument so $\phi_1$ corresponds to a path from zero to the root with argument closest to $\pi$. 

Appealing to the Newton-Puiseux theorem \cite{ebeling2007functions}, we can write $F_\varepsilon$ as a fractional power series on each irreducible component of the polar curve. In our case there is only one possible leading order exponent given by\footnote{This can be calculated in numerous ways. In our case, the polar curve and singularity are simple enough to write down the Puiseux expansion explicitly. Alternatively, one can examine the Newton polygon of the singularity \cite{willis2008compute} or implement the {\tt AsymptoticSolve[$\cdot$]} function in Mathematica.}
\begin{gather}
    \beta=\frac{nk}{n-1}.
\end{gather}
As such, we can take the radial function $\psi(t)=\beta-1$ everywhere and the collection of sets $T_m$ are given by 
\begin{gather}\label{eqn:tm}
    T_m=\{t:\arg t\geq\pi(1+2(m-\beta))\}. 
\end{gather}
We have that $T_m=\mathbb{C}$ for $m\leq \beta-1$ and that $T_m=\varnothing$ for $m\geq\beta$ but if $\beta-1<m<\beta$, the set $T_m$ will be neither empty nor all of $\mathbb{C}$.
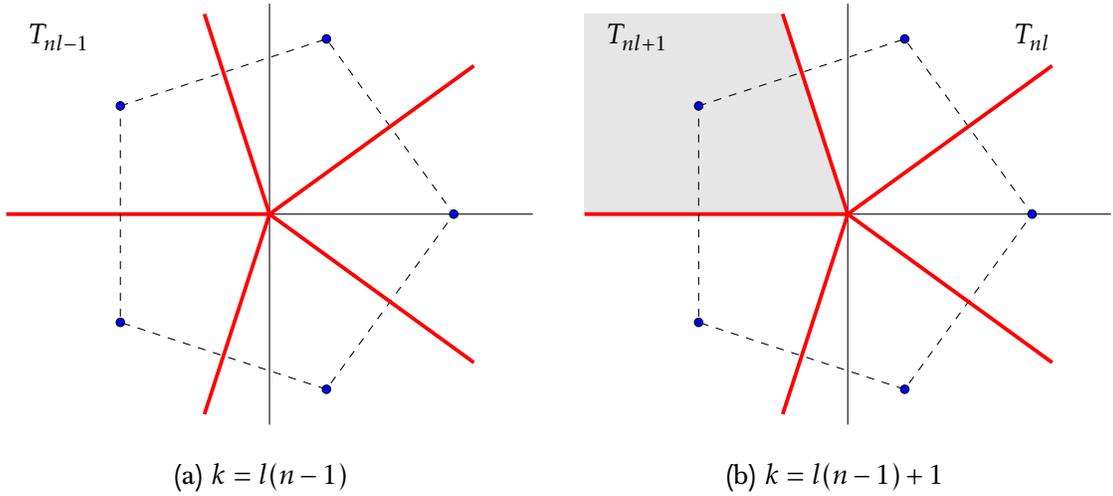
\begin{figure}[]
    \begin{center}	
        \begin{tikzpicture}[scale=0.7]		
            \draw[color=red, line width=0.5mm] (-5, 0) -- (0, 0);
            \draw (0,0) -- (5,0);		
            \draw (0,-4) -- (0,4);		
            \draw[dashed] (-2.832, 2.057) -- (1.082, 3.329) -- (3.5, 0) -- (1.082, -3.329) -- (-2.832, -2.057) -- ((-2.832, 2.057);		
            \draw[fill=blue] (-2.832, 2.057) circle(0.08);		
            \draw[fill=blue] (1.082, 3.329) circle(0.08);		
            \draw[fill=blue] (3.5, 0) circle(0.08);		
            \draw[fill=blue] (1.082, -3.329) circle(0.08);		
            \draw[fill=blue] (-2.832, -2.057) circle(0.08);		
            \draw[color=red, line width=0.5mm] (0, 0) -- (-1.236, 3.804);		\draw[color=red, line width=0.5mm] (0,0) -- (3.883, 2.821);		
            \draw[color=red, line width=0.5mm] (0,0) -- (3.883, -2.821);		\draw[color=red, line width=0.5mm] (0,0) -- (-1.236, -3.804);		
            \draw (-4,3.8) node[anchor=north]{$T_{nl-1}$};		
            \draw (-2,-5) node[anchor=west]{(a) $k=l(n-1)$};	
        \end{tikzpicture}
\;\;\;\;		
        \begin{tikzpicture}[scale=0.7]
            \draw[fill=gray!20, color=gray!20] (0, 0) -- (-5, 0) --(-5, 3.804)--(-1.236, 3.804) -- (0,0);		
            \draw[color=red, line width=0.5mm] (-5, 0) -- (0, 0);		
            \draw (0,0) -- (5,0);		
            \draw (0,-4) -- (0,4);		
            \draw[dashed] (-2.832, 2.057) -- (1.082, 3.329) -- (3.5, 0) -- (1.082, -3.329) -- (-2.832, -2.057) -- ((-2.832, 2.057);		
            \draw[fill=blue] (-2.832, 2.057) circle(0.08);		
            \draw[fill=blue] (1.082, 3.329) circle(0.08);		
            \draw[fill=blue] (3.5, 0) circle(0.08);		
            \draw[fill=blue] (1.082, -3.329) circle(0.08);		
            \draw[fill=blue] (-2.832, -2.057) circle(0.08);		
            \draw[color=red, line width=0.5mm] (0, 0) -- (-1.236, 3.804);		\draw[color=red, line width=0.5mm] (0,0) -- (3.883, 2.821);		
            \draw[color=red, line width=0.5mm] (0,0) -- (3.883, -2.821);		\draw[color=red, line width=0.5mm] (0,0) -- (-1.236, -3.804);		
            \draw (-4,3.8) node[anchor=north]{$T_{nl+1}$};		
            \draw (3.5,3.8) node[anchor=north]{$T_{nl}$};		
            \draw (-2.5,-5) node[anchor=west]{(b) $k=l(n-1)+1$};	
        \end{tikzpicture}
    \end{center}
\caption{A visualisation of how the covering of $\mathbb{C}$ by the sets $T_m$ changes under $k\mapsto k+1$. Specifically we see that $m\mapsto m+1$ for all but exactly one critical point, where we get $m\mapsto m+2$. Here we have taken $n=6$ and each critical point of $f|_{z=\varepsilon}$ is denoted by a blue dot.}\label{fig:tm}
\end{figure}

First note that if $(n-1)|k$ then $(\beta-1)\in \mathbb{N}$ and $M_i=\max\{m:\varphi_i\in T_m\}=kn/(n-1)-1$ for $i=1,\ldots,n-1$. Now for convenience, let $k=l(n-1)$ and consider what happens to $\beta$ under the map $k\mapsto k+1$,
\begin{gather*}
    \beta=nl\mapsto nl+\frac{n}{n-1}=nl+1+\frac{1}{n-1}. 
\end{gather*}
From this it is clear that $M_i\mapsto M_i+1=nl$ for $i\neq 1$, but we also have $M_1\mapsto M_1+2=nl+1$. This can be seen from the fact that
\begin{gather}
    T_{nl+1}=\bigg\{t:\arg t \geq \pi\Big(1-\frac{2}{n-1}\Big)\bigg\}.
\end{gather}
Since the critical values of $F_\varepsilon$ form a regular $(n-1)$-gon around the origin, it is guaranteed that exactly one root will lie in this range, as shown in Figure \ref{fig:tm}. By repeatedly increasing $k$ by increments of one, and noting that the base case of $k=1$ is simply the $A_1$ diagram, we can inductively show that 
\begin{gather}
    M_i=k+\bigg\lfloor \frac{k}{n-1} \bigg\rfloor -\Theta(i-\mathrm{mod}(k,n-1)),
\end{gather}
where $\Theta(\cdot)$ is a Heaviside step-function. That is, $M_i=k+\big\lfloor\tfrac{k}{n-1}\big\rfloor$ if $i\leq \mathrm{mod}(k,n-1)$ and $M_i=k+\big\lfloor\tfrac{k}{n-1}\big\rfloor-1$ otherwise.

With this, and theorem \ref{th:gab}, we now have a simple method to write down the Coxeter-Dynkin diagram. Let $N_1=k+\big\lfloor\tfrac{k}{n-1}\big\rfloor-1$ and $N_2=\mathrm{mod}(k,n-1)$. Then the Coxeter-Dynkin diagram will have the form
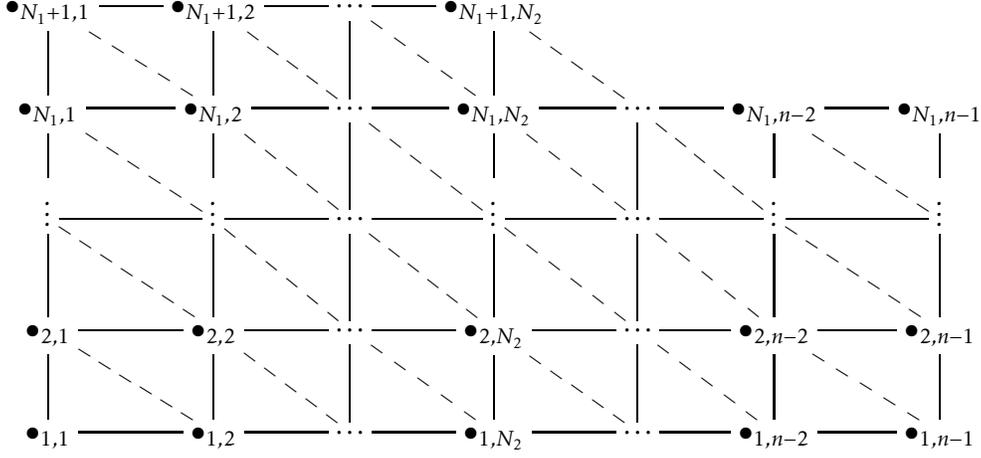
\begin{gather*}
    \xymatrix{
        \bullet_{N_1+1,1} \ar@{-}[r]\ar@{-}[d]\ar@{--}[dr] &
        \bullet_{N_1+1,2} \ar@{-}[r]\ar@{-}[d]\ar@{--}[dr] &
        \cdots \ar@{-}[r]\ar@{-}[d]\ar@{--}[dr] &
        \bullet_{N_1+1, N_2} \ar@{-}[d]\ar@{--}[dr]
    \\
        \bullet_{N_1,1}\ar@{-}[r]\ar@{-}[d]\ar@{--}[dr]& \bullet_{N_1,2}\ar@{-}[r]\ar@{-}[d]\ar@{--}[dr]& \cdots \ar@{-}[r]\ar@{-}[d]\ar@{--}[dr] & \bullet_{N_1,N_2} \ar@{-}[r]\ar@{-}[d]\ar@{--}[dr] & \cdots \ar@{-}[r]\ar@{-}[d]\ar@{--}[dr] & \bullet_{N_1,n-2}\ar@{-}[r]\ar@{-}[d]\ar@{--}[dr]& \bullet_{N_1, n-1} \ar@{-}[d]
    \\
        \vdots\ar@{-}[r]\ar@{-}[d]\ar@{--}[dr] & \vdots\ar@{-}[r]\ar@{-}[d]\ar@{--}[dr] &\cdots\ar@{-}[d]\ar@{-}[r]\ar@{--}[dr] & \vdots\ar@{-}[r]\ar@{-}[d]\ar@{--}[dr] &\cdots\ar@{-}[d] \ar@{-}[r]\ar@{--}[dr]& \vdots\ar@{-}[r]\ar@{-}[d]\ar@{--}[dr] & \vdots\ar@{-}[d]
    \\
        \bullet_{2,1}\ar@{-}[d]\ar@{--}[dr] \ar@{-}[r] & \bullet_{2,2}\ar@{-}[d] \ar@{--}[dr]\ar@{-}[r]&\cdots\ar@{-}[d]\ar@{--}[dr]\ar@{-}[r]& \bullet_{2, N_2}\ar@{-}[d]\ar@{-}[r]\ar@{--}[dr] &\cdots\ar@{-}[d]\ar@{-}[r]\ar@{--}[dr]  & \bullet_{2,n-2}\ar@{-}[d]\ar@{--}[dr]\ar@{-}[r]&\bullet_{2,n-1} \ar@{-}[d]
    \\
        \bullet_{1,1} \ar@{-}[r] & \bullet_{1,2}\ar@{-}[r]&\cdots\ar@{-}[r]&
        \bullet_{1, N_2}\ar@{-}[r] & \cdots\ar@{-}[r]  &
        \bullet_{1,n-2}\ar@{-}[r] & \bullet_{1,n-1}
    }
\end{gather*}

From the Coxeter-Dynkin diagram of a singularity, we can construct the associated quiver using the methods described in \cite{ringel2006tame, Cecotti:2011gu, 2010arXiv1001.1531K}. Loosely speaking, for a theory $\mathcal{T}$ one chooses an orientation for the solid arrows to form the associated quiver $\hat{\mathfrak{Q}}(\mathcal{T})$. The dashed arrows are then interpreted as relations imposed on the representation category $\mathbf{Rep}(\hat{\mathfrak{Q}}(\mathcal{T}))$. This constrains the possible orientations so we are left with relatively few choices. Finally, in the auxiliary supersymmetric quantum mechanics defined by the quiver, the relations imposed from the dashed arrows arise from F-terms in the superpotential. We therefore add an appropriate Lagrange multiplier to the superpotential which imposes this commutativity. This forms the 3-Calabi-Yau completion of the quiver by taking each dashed arrow and replacing it with a solid arrow in the opposite direction. We call the quiver obtained from a Coxeter-Dynkin diagram together with 3-Calabi-Yau completion in this way $\mathfrak{Q}(\mathcal{T})$.

Since the diagrams are composed of partially completed squares we can form the quiver $\mathfrak{Q}(\mathcal{T})$ by composing orientations of squares. We therefore specify the orientation of the quivers by taking
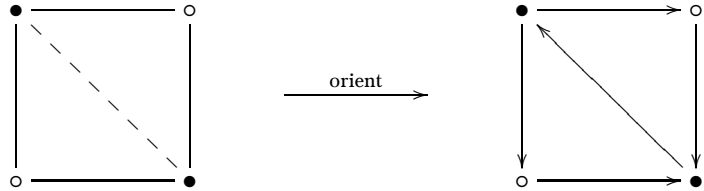
\begin{gather*}
    \xymatrix{
    \bullet \ar@{-}[rr]\ar@{-}[dd] \ar@{--}[ddrr] & & \circ \ar@{-}[dd] & & & & \bullet \ar[rr]\ar[dd] & & \circ \ar[dd]\\
    & & & \ar[rr]^{\text{orient}} & & \\
    \circ \ar@{-}[rr] & & \bullet & & & & \circ \ar[rr] & & \bullet \ar[uull]
    }
\end{gather*}
so we take the arrows going from top-left to bottom-right around a dashed line. From this, we can extrapolate this orientation to all diagrams. Notice that the additional line from taking the 3-CY completion gives additional non-trivial loops which can be included in the superpotential for the quiver quantum mechanics.

Having derived the general form of the $A_{n-1}^{(n-1)}[k]$ quivers let us check that they coincide with known results. Recall that the Milnor number of the singularity corresponds to the number of nodes in the quiver. These numbers can be calculated using (\ref{MN}). The total number of nodes in the quiver is given by
\begin{gather}
    \mu(A_{n-1}^{(n-1)}[k])=\sum_{i=1}^{n-1}M_i=(n-1)\bigg(k+\bigg\lfloor\frac{k}{n-1}\bigg\rfloor\bigg)-(n-1-\mathrm{mod}(k,n-1)).
\end{gather}
We can rewrite the floor function to give the expression
\begin{gather}
     \mu(A_{n-1}^{(n-1)}[k])=(n-1)\bigg(k-1+\frac{k-\mathrm{mod}(k,n-1)}{n-1}\bigg)+\mathrm{mod}(k,n-1)=nk-(n-1),
\end{gather}
in agreement with the result of Xie and Wang \cite{Wang:2015mra}.

Inputting our quivers in the Mathematica and computing the cokernels for $3 \leq n \leq 15$ and $2\leq k \leq 15$ we see total agreement with the results of the previous section. That is, $\mathrm{Tor}\,\mathbb{D}$ is trivial.

\subsection{Extension to $E_6$ and $E_8$ cases}
Having dealt with the $A_{n-1}^{(n-1)}[k]$ case, the $E_6^{(b)}[k]$ and $E^{(b)}_8[k]$ cases follow easily from the following observation.

Notice that in these cases the singularities are stably equivalent to 
\begin{gather}
    x^{i+1} + y^{j+1} + y z^k =0,
\end{gather}
up to a possible relabelling of $x \leftrightarrow y$ in comparison to Table \ref{tbl:cdv}. As the $x$ terms do not mix with any other variable, we have that singularity is a direct sum of the $x$ singularity and the singularity defined by $y^{j+1}+y z^k=0$. We analysed this singularity in the previous section so, in terms the BPS quiver representative $\mathfrak{Q}(\mathcal{T})$ obtained from the Coxeter-Dynkin diagram, we propose that
\begin{gather}\label{eqn:boxprod}
    \mathfrak{Q}(E_{n}^{(b)}[k])=A_{i}\;\boxtimes\; \mathfrak{Q}(A_{j}^{(j)}[k]),
\end{gather}
where $i$ and $j$ are inferred from Table \ref{tbl:cdv}. We list all of the quivers obtained in this way in Table \ref{tbl:direct-sums}.
\begin{table}[h]
\centering
\begin{tabular}{|c|c|c|}
\hline
\rowcolor[HTML]{DAE8FC} 
\textbf{Theory $\mathcal{T}$} & \textbf{Quiver $\mathfrak{Q}(\mathcal{T})$}    & \textbf{Milnor number}\\ \hline
$E_6^{(8)}[k]$                & $A_3 \; \boxtimes \; \mathfrak{Q}(A_2^{(2)}[k])$ & $3(3k-2)$\\
$E_6^{(9)}[k]$                & $A_2 \; \boxtimes \; \mathfrak{Q}(A_3^{(3)}[k])$ & $2(4k-3)$\\
$E_8^{(20)}[k]$               & $A_4 \; \boxtimes \; \mathfrak{Q}(A_2^{(2)}[k])$ & $4(3k-2)$\\
$E_8^{(24)}[k]$               & $A_2 \; \boxtimes \; \mathfrak{Q}(A_4^{(4)}[k])$ & $2(5k-4)$\\ \hline
\end{tabular}
\caption{Direct sum structure of $E_n^{(b)}[k]$ theories for $n=6$ and $n=8$.}\label{tbl:direct-sums}
\end{table}

Using this decomposition we can easily calculate the cokernels from which we see total agreement with the Orlik's algorithm result. This heavily suggests that a possible BPS quiver for these theories is indeed given by \eqref{eqn:boxprod}. 

The polar curve method can also be used to find the BPS quivers for the $D_n$ and $E_7$ cases, but identifying the correct $f|_{z=0}$ Coxeter-Dynkin diagram is more subtle. We hope to return to this in a later paper focused fully on BPS quivers.

\section*{Acknowledgements}
We would like to thank Michele Del Zotto and Iñaki García Etxebarria for the interesting discussions and helpful suggestions about the manuscript, as well as
initial inspiration for this paper. We would also like to thank Viktor Matyas for many useful comments. S.S.H. is funded by the STFC grant ST/T506035/1. The work of R.M. is in the context of the ERC project MEMO, which is funded by the European Research Council (ERC) under the European Union’s Horizon 2020 research and innovation programme (grant agreement No. 851931). 

\appendix

\section{A quick review of the polar curve method}\label{ap:polar}
For completeness, we will give a quick overview of the polar curve method. Our discussion will mostly follow \cite{arnol1981singularity}.

Let $f:(\mathbb{C}^n, 0)\rightarrow (\mathbb{C},0)$ be a function germ corresponding to an isolated singularity at 0 and let $z:(\mathbb{C}^n,0)\rightarrow (\mathbb{C},0)$ be a linear function. The {\it polar curve} $\gamma_z(f)$ of $f$ relative to $z$ is the set of critical points of the map $(f,z):(\mathbb{C}^n,0)\rightarrow (\mathbb{C}^2, 0)$. In practice, this is simply the union of critical points of $f-\lambda z$ with $\lambda$ varying.

Assuming $z$ is linear in position, as will be the case for us, the restriction $f|_{z=0}$ will also have an isolated singularity at 0. Furthermore, as the singularity in $f|_{z=0}$ is isolated it follows that none of the irreducible components $\gamma_i$ of $\gamma_z(f)$ are contained in the hyperplane $\{z=0\}$ and that $f|_{\gamma_i}\neq 0$ on each irreducible component. 

The aim is to now study the behaviour of the $f|_{\gamma_i}$ as $z\rightarrow 0$. The Newton-Puiseux theorem \cite{ebeling2007functions} states that on each branch of an algebraic curve $\Xi(x,y)=0$ there exists a solution $y=\xi(x)$ that is given by a fractional power series which we call the Puiseux expansion on that component/branch. We therefore Puiseux expand the $f|_{\gamma_i}$ as
\begin{gather}
    f|_{\gamma_i}\sim c_i z^{\beta_i},\;\; z\rightarrow 0,
\end{gather}
to understand the small $z$ regime. Furthermore, the critical points of $f-\lambda z$ lying on the polar curve components are determined by
\begin{gather}
    \partial_z f|_{\gamma_i}=\lambda \Rightarrow \lim_{\lambda\rightarrow 0}(c_i \beta_i z^{\beta_i-1}+\ldots)=0.
\end{gather} 
Hence, $\beta_i>1$. The Puiseux exponents help characterise the behaviour of vanishing cycles of $f$. As such, we now need to define some objects to help this characterisation.

Consider the deformed function $F_\varepsilon=f|_{z=\varepsilon}$. The asymptotic expansion of $F_\varepsilon$ near critical values is given by
\begin{gather}
    F_\varepsilon\sim c_i \varepsilon^{\beta_i},
\end{gather}
where $\beta_i$ is the Puiseux exponent for the irreducible component $\gamma_i$ where the critical value lies. For each such exponent, let $\mathbf{A}_{\beta_i}$ be the annulus that contains all critical values of $F_\varepsilon$ with Puiseux exponent $\beta_i$. For small $\varepsilon$, these annuli can be chosen to be disjoint. We now wish to define a non-increasing radial function $\psi(t)$ such that $\psi(t)=\beta_i-1$ on $\mathbf{A}_{\beta_i}$. The collection of sets 
\begin{gather}
    T_m=\{t:\arg t \geq \pi(2m-1-2\psi(t))\},
\end{gather}
have the useful property that $\mathbf{A}_{\beta}\subset T_m$ for $m\leq \beta-1$ and $T_m \cap \mathbf{A}_\beta=\varnothing$ for $m\geq \beta$. Recall that we can define a distinguished basis of vanishing cycles by specifying a distinguished system of paths connecting 0 to the critical values of $F_\varepsilon$. Such a system of paths is said to be admissible if each path $\varphi_i:[0,1]\rightarrow \mathbb{C}$ is contained within the set $T_m$ with $\varphi(1)\in T_m$. 

The polar curve method uses an admissible system of paths to relate the intersection matrices of $f$ and $f|_{z=0}$. Specifically, we have the following theorem.
\begin{theorem}\label{th:gab}
    {\bf (Gabrielov \cite{gabrielov}).} Let $f$ be an isolated singularity depending on a linear function $z$ and $\{\varphi_i : i=1,\ldots, \mu(f|_{z=0})\}$ be an admissible system of distinguished paths connecting 0 to the critical values of $f|_{z=\varepsilon}$. There exists a distinguished basis of vanishing cycles $\Delta_{m,i}$ with $m=1,\ldots, M_i$, where $M_i=\max\{m:\varphi_i\in T_m\}$,\footnote{Here we have imposed the condition that $\varepsilon$ is chosen such that no critical points lie on the boundary of any $T_m$ or the negative real half-line $\mathbb{R}_-$ so the upper limit of $m$ is well defined.} such that
    \begin{gather*}
        \Delta_{m,i} \circ \Delta_{m,j} =\Delta_i \circ \Delta_j, \;\;\;\;\; \Delta_{m,i} \circ \Delta_{k,i}=1\; \; \text{if } |m-k|=1,\\
        \Delta_{m,i} \circ \Delta_{k,j}=-\Delta_i \circ \Delta_j \;\; \text{if } |m-k|=1\;\text{with } (m-k)(i-j)<0,
    \end{gather*}
    and
    \begin{gather*}
        \Delta_{m,i}\circ\Delta_{k,j}=0,
    \end{gather*}
    if $|m-k|>1$ or $|m-k|=1$ with $(m-k)(i-j)>0$. Here $\{\Delta_i\}$ is a distinguished basis of vanishing cycles for $f|_{z=0}$.
\end{theorem}
All in all, the polar curve method takes a singularity and identifies a reduced singularity from which one can infer the whole intersection matrix by investigating an admissible system of paths and the sets $T_m$.

\addcontentsline{toc}{section}{References}
\bibliography{main.bib}{}
\end{document}